# Low temperature anomaly of heat capacity of $CD_4$ rotors in solid $CD_4$-Kr solution.

M.I. Bagatskii, V.V. Dudkin, V.G. Manzhelii, D.A. Mashchenko, S.B. Feodosiev.

*Verkin Institute for Low Temperature Physics and Engineering, National Academy of Sciences of Ukraine, 47, Lenin Ave., 61103, Kharkov, Ukraine.*

*e-mail: dudkin@ilt.kharkov.ua*

*The heat capacity of the solid $Kr$-$CD_4$ (13% $CD_4$) solution has been investigated. It is shown that the temperature dependence of the heat capacity $C_{rot}$ of the rotational subsystem in this solution is radically different from the corresponding dependences in the previously studied $Kr$-$CD_4$ (1%, 5% $CD_4$) and $Kr$-$CH_4$ (5-60%) solutions. A model is proposed to explain the observed dependence $C_{rot}(T)$. The experimental results can be described taking into account the contribution to the heat capacity from $CD_4$ molecules that are in weak, medium and strong molecular fields. The mean concentrations and energy differences between the ground and first excited energy levels of the $CD_4$ molecules in these molecular fields have been estimated.*

Solid solutions of methane and deuteromethane in krypton are suitable objects for investigating the effect of zero (quantum) rotation of rotors and the rotor interaction on the dynamics of an ensemble of rotors. The rotational constant $B$ ($B = \hbar^2/2I$, where $\hbar$ is the Planck constant, $I$ is the moment of the inertia of the molecule) of the $CH_4$ molecule is two times larger than that of the $CD_4$ molecule. It is possible to separate the contributions of matrix-isolated molecules and molecular clusters of different sizes and configurations to the heat capacity of solutions by varying the concentration of rotors. This will also provide information about the interaction between the rotors in the lattice. Through a comparison of the heat capacities of solutions of Kr with $CH_4$ and $CD_4$ we can better understand the role of the parameters changing with isotopic substitution of atoms and the quantum effects (more pronounced in the $CH_4$-Kr system) in the rotational motion of molecules. In this connection, it is important to note that the $CH_4$ and $CD_4$ molecules can exist in the form of three nuclear spin species – E, T, A, the total nuclear spins being 0, 1, and 2, and 0, 2, 4 respectively. The energy spectra species are different, so the heat capacity of the rotor ensemble is essentially dependent on nuclear spin species concentrations and conversion rate.

Dilute solutions with up to 10% $CH_4$, and up to 5% $CD_4$ in Kr were investigated using calorimetric [1-4] and neutron diffraction [5] methods. It was found [1-5] that the rotation of the rotors in isolated one-molecule and two-molecule clusters is weakly hindered. The low-energy part of the spectrum of isolated $CD_4$ molecules is similar to reduced (approximately by a factor of 2.5) one of $CH_4$ molecules. The energy differences $E_{AT}$ between the lowest levels of the nuclear spin species A and T in the spectrum of the isolated $CH_4$ and $CD_4$ rotors in the solutions are respectively 0.77 and 0.62 times smaller then the energy differences of free rotors (this relation shows the degree of the hindering of the rotation). The low-temperature calorimetric investigation of $CH_4$-Kr solutions with $CH_4$ concentrations up to 60% [4] shows that the character of $CH_4$ molecules rotation at low temperatures remains qualitatively invariable up to 60% $CH_4$. As the $CH_4$ concentration grows from 5% to 60%, the rotation of the $CH_4$ molecules becomes increasingly hindered and $E_{AT}$ changes from 11.7 to 7.4 K, respectively. The similar qualitative behavior of rotors was expected in $CD_4$-Kr solutions with $CD_4$ concentrations over 5%. However, the heat capacity measurement shows that the motion of the rotors in the 13% $CD_4$-Kr solution is in a sharp contrast to what is observed in the solutions with up to 60% $CH_4$ or 5% $CD_4$.

The heat capacity of the solid 13% $CD_4$-Kr solution was measured using an adiabatic calorimeter [6] at $T$=0.6–20 K. The time of calorimetric heating $t_h$ was 2-4 minutes. The effective time $t_m$ of a single heat capacity measurement was $t_m=t_h+t_e$, where $t_e$ is the time required to achieve a steady-state temperature behavior from the time the heating was turned off. $t_e$ varied within ~ 40–10 min in the interval 0.6–20 K. The compositions of the used gases were: $CD_4$ – of isotopic purity 99%, the purity of the other gases 99.20% ($N_2$ – 0.5%; $O_2$ – 0.2%, CO – 0.1% and Ar < 0.01%); the purity of Kr was 99.79% (Xe – 0.2%, $N_2$ – 0.01%, $O_2$ and Ar < 0.01%). The solid sample was prepared from the gas phase at $T \approx 82$ K, which provided a random distribution of the $CD_4$ molecules and homogeneity of the sample. Before the experiment the sample was kept at the lowest temperature of measurement for about 24 hours. The temperature relaxation of the calorimeter after fast heating or cooling shows that the nuclear spin conversion is a relatively rapid process. In all our measurements the distribution of the nuclear spin species was close to equilibrium. The error of the heat capacity measurement was 6% at 0.7 K, 2% at 1 K, 1% at 2 K and 0.5% above 4 K.

The contribution of the rotational subsystem was found by subtracting the heat capacity of the lattice from the total heat capacity of the solution. The lattice heat capacity was calculated as in [2]. Figure 1 shows the temperature dependence of the contribution of the rotors to the heat capacity normalized to the $CD_4$ concentration $n$ and to the universal gas constant $R$: $C_R(T)=(C-C_{lat})/(nR)$. It is very different from the dependences measured on $CD_4$-Kr solutions with lower $CD_4$ concentration (1%, 5%) [1, 2] and $CH_4$-Kr solutions with 5-60% $CH_4$ [3, 4].

The most prominent feature of $C_R(T)$ is its maximum near 1 K in the Kr-13%CD$_4$ solution. A similar maximum of heat capacity is observed if several closely-spaced lowest energy levels in the energy spectrum are separated from the higher levels with the relatively large gap. In the following our analysis of the results will be confined to the interval 0.6–2 K. In this temperature regions $C_R(T)$ is mainly determined by the lowest energy levels of the A and T species, which we can estimate (see below). At the same time, information about the higher energy levels responsible for $C_R(T)$ at T>2 K is not available and no reliable data can be obtained from our measurement. The temperature of the $C_R(T)$ maximum measured experimentally can be obtained using the two-level model at $E_{AT} \approx 3$ K. This value is much below $E_{AT} = 4.8$ K [5] for the matrix-isolated molecules in the Kr-1%CD$_4$ solution. This implies that the spectrum of a part of the rotors in the solution is appreciably deformed in comparison with the spectrum of the isolated molecules. Meanwhile, the absolute values of $C_R(T)$ at 0.6-2 K obtained within the two-level model nearly twice exceed the corresponding experimental results. This is possible only when a considerably part of the rotors has spectra with $E_{AT} < 1$ K and the contribution of these rotors to $C_R(T)$ in this temperature region is negligible.

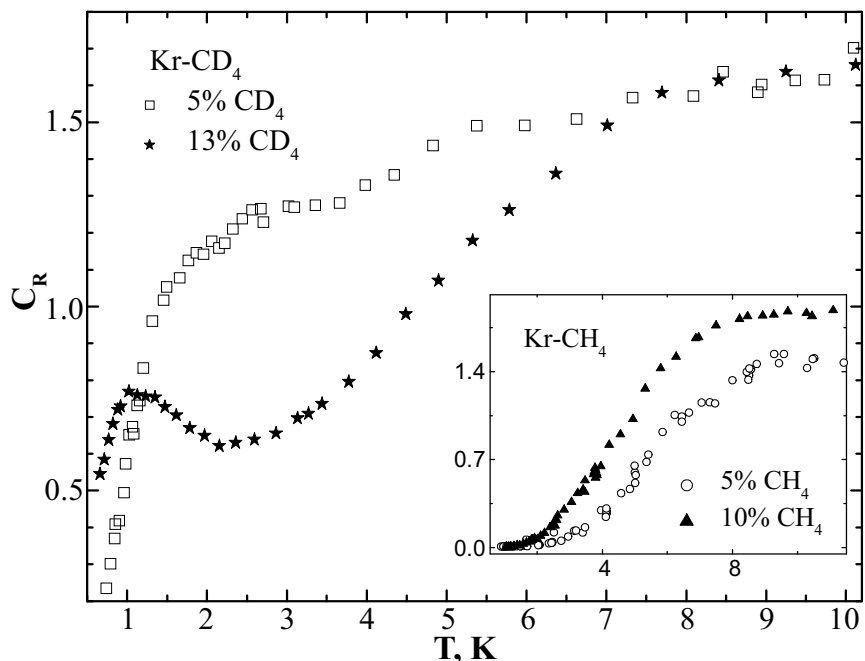

Fig 1. Normalized heat capacity $C_{rot}/(Rn)$ of $(CD_4)_n Kr_{1-n}$ and $(CH_4)_n Kr_{1-n}$ solid solutions.

Note that in the Kr-13%CD$_4$ solution a part of the rotors $n_s = (1-n)^{12} = 0.19$ are matrix-isolated and we can expect that some of them are in a weak field with $E_{AT}=4.8$ K [5] (see below).

The rotational spectrum of the molecule is determined by the symmetry and the noncentral (molecular and crystal) field experienced by the molecule. As the noncentral field increases, the low-energy part of the spectrum of one molecule changes as follows: the lowest levels of the nuclear-spin species go lower and draw closer. In strong fields they form a group of tunnel levels separated by a considerable gap from the next group of levels [7,8]. In a growing noncentral field the rotational motion of the molecules varies from weakly hindered to strongly hindered and further to orientational vibrations (librations). Proceeding from the aforesaid, we can describe the rotational subsystem of the solution introducing a concept of three groups of rotors with $E_{AT} = 4.8$ K, $E_{AT} \approx 3$ K and $E_{AT} < 1$ K. We calculated the heat capacity using the energy spectra (Fig. 2) corresponding to the systematics of the spectra of $CD_4$ rotation in Kr [8,9]. Hereinafter, the rotors that are in weak, medium and strong noncentral fields are specified as WH-, H- and SH-rotors, respectively. The best agreement with experiment was obtained (Fig. 2) using the parameters of Tab. 1.

Table 1. Parameters obtained from analysis of experimental $C_R$ results $C_{rot}(T)$ for the Kr-13%$CD_4$ solution (weak-medium-strong field model).

| molecular field | motion | relative molecular concentration | $E_{AT}$, K |
|---|---|---|---|
| weak | weakly hindered rotation | $n_{WH} = 0.1$ | 4.8 |
| medium | hindered rotation | $n_H = 0.47$ | 2.64 |
| strong | strongly hindered rotation | $n_{SH} = 0.43$ | 0.5 |

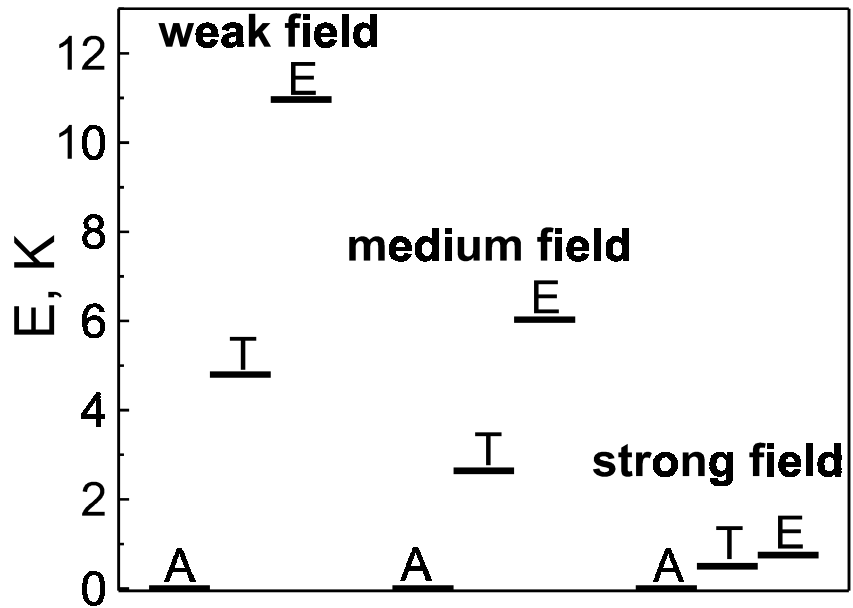

Fig 2. Spectra of rotors in effective weak, medium and strong fields obtained from the analysis of $C_R(T)$ of Kr-13% $CD_4$ solid solution.

As shown in [2], the solution with 1% and 5% $CD_4$ have practically equal effective gaps $E_{AT}$. Hence, the spectra of isolated $CD_4$ pairs and isolated single molecules are very close. The sharp change in $E_{AT}$ of the 13% $CD_4$-Kr solution may be attributed to the increased contents of $CD_4$ rotors that have two or more $CD_4$ rotors among the nearest neighbors. In this solution their contents amount 0.47. This is close to the portion of SH-rotors $n_{SH} = 0.43$. The relative concentration of the WH-rotors $n_{WH} = 0.1$ is noticeably lower than the concentration of isolated rotors (0.19). The difference can be explained qualitatively assuming that the mean molecular field excited by the molecules is strongly dependent on the character of zero-rotation motion of the $CD_4$ and $CH_4$ molecules in the solid $CD_4$-Kr and $CH_4$-Kr solution. The SH-rotors therefore excite a very strong molecular field whose influence is significant even outside the first coordination sphere. In this case the spectrum of isolated molecules and pairs of molecules containing SH-rotors in the second coordination sphere can be deformed considerably, and the rotation of such molecules is no longer weakly hindered.

Note that because of the distinctions in the local symmetry of the configurations and in the number of rotors among the nearest neighbors, the energies shown here are effective values for each of the above groups of rotors. On the whole, the model provides a fairly good description of the heat capacity in the Kr-13%$CD_4$ solution and agrees qualitatively with the notion of the behavior of the system of rotors with a stronger noncentral interaction than that between the rotors. Indeed, because the rotational constants and the character of the zero-rotation motion are much different for the $CD_4$ and $CH_4$ molecules in solid Kr, the noncentral interaction of $CD_4$ molecules is significantly stronger than that of $CH_4$ molecules.

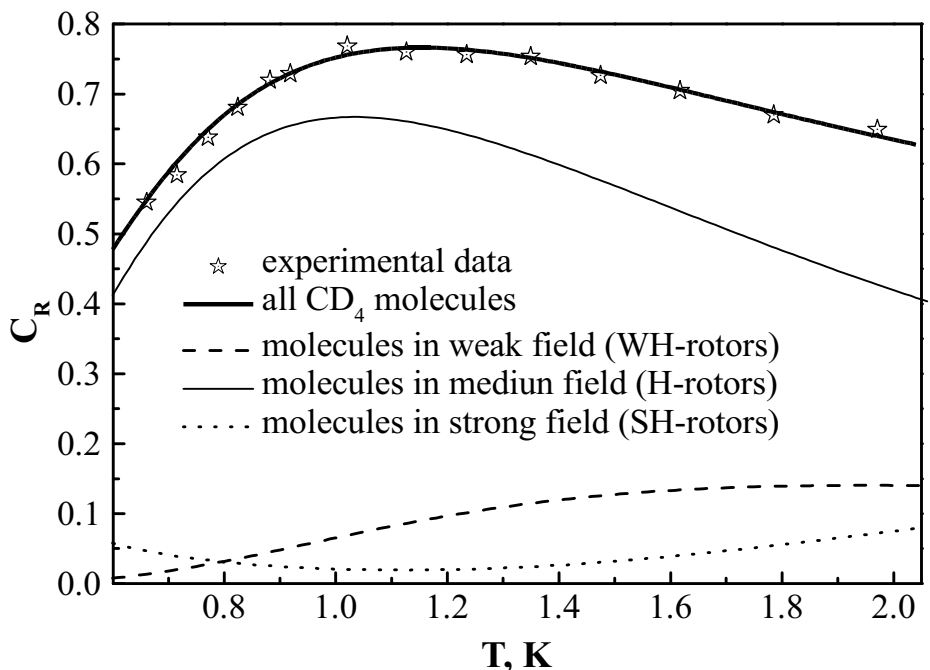

Fig 3. Temperature dependence of $C_R$ of Kr-13%$CD_4$ solid solution. The lines show contributions of rotors having different types spectra (weak-medium-strong field model).

The work was supported by the Ukraine Minister of Education and Science (Project F 7/286/3-2004).

REFERENCES

1. M.I. Bagatskii, V.G. Manzhelii, D.A. Mashchenko, V.V. Dudkin, *Low Temp. Phys.* **29,** 159 (2003).
2. M.I. Bagatskii, V.G. Manzhelii, D.A. Mashchenko, V.V. Dudkin, *Low Temp. Phys.* **29,** 1028 (2003).
3. I.Ya. Minchina, V.G. Manzhelii, M.I. Bagatskii, O.V. Sklyar, D.A. Mashchenko, and M.A. Pokhodenko, *Low Temp. Phys.* **27**, 568 (2001).
4. M.I. Bagatskii, V.G. Manzhelii, I.Ya. Minchina, D.A. Mashchenko, I.A. Gospodarev, J. Low Temp. Phys. **127**, 459 (2003).
5. B. Asmussen, W. Press, M. Prager, and H. Blank, *J. Chem. Phys.* **98**, 158 (1993).
6. M.I. Bagatskii, I.Ya. Minchina, V.G. Manzhelii, *Sov. J. Low Temp. Phys.* **10**, 542 (1984),
7. T. Yamamoto, Y. Kataoka, K. Okada, *J. Chem. Phys.* **66**, 2701 (1977).
8. A. Hüller, J. Raich, *J. Chem. Phys.* **71**, 3851 (1979).
9. K. Nishiyama, *J. Chem. Phys.* **56**(10), 5096 (1972).